**Collective Nucleation Dynamics in Two-dimensional Emulsions with Hexagonal Packing**

**Samira Abedi, Chau-Chyun Chen\* and Siva A. Vanapalli\***

a Department of Chemical Engineering, Texas Tech University, Lubbock, TX 79409-3121, USA

\*Corresponding author. Email address: siva.vanapalli@ttu.edu; chauchyun.chen@ttu.edu

**Abstract**

We report a new mechanism for nucleation in a monolayer of hexagonally packed monodisperse droplet arrays. Upon cooling, we observe solidified droplets to nucleate their supercooled neighbors giving rise to an autocatalytic-like mechanism for accelerated crystallization. This collective mode of nucleation depends on the strength and nature of droplet contacts. Intriguingly, the statistical distribution of the solidified droplet clusters is found to be independent of emulsion characteristics except surfactant. In contrast to classical nucleation theory, our work highlights the need to consider collective effects of nucleation in supercooled concentrated emulsions where droplet crowding is inevitable.

Emulsions are manipulated to form a wide range of soft materials including low-viscosity fluids, gels, elastic pastes and glasses [1-3]. This richness in functionality is due to the ability to fine tune the physicochemical properties of the individual phases as well as the interface [4-7]. A common approach to manipulate functionality is to use crystallizable oils as the dispersed phase and cool the emulsions so that the nucleated droplets form partially-coalesced networks [8, 9], imparting unique rheological properties [10, 11]. More recently, this thermal quench has become an attractive route to engineer novel emulsions where droplets have a non-spherical shape [12-14] or the capacity to self-shape [15].

During the thermal quench, droplets in the emulsion nucleate undergoing a liquid-solid phase transformation. The mechanisms of nucleation in emulsions have been long-studied [16, 17]. The simple picture is that when the material to be dispersed is divided into droplets, only a fraction has impurities, necessitating significant undercooling to induce nucleation in the impurity-free droplets. Thus, majority of the droplets undergo homogeneous nucleation in which the crystal nucleus formed due to local density fluctuations can grow, while a small fraction undergoes heterogeneous nucleation.

It is now well recognized that the above simple picture is insufficient to explain observed rates of nucleation in emulsions [18, 19]. Studies show that nucleation rates depend on droplet size with larger droplets requiring smaller undercooling [20] suggesting that polydispersity can confound results [21, 22]. Nucleation rates can also be sensitive to surfactant type since these interfacial impurities can promote heterogeneous nucleation [23, 24]. Strikingly, addition of solidified droplets to a supercooled emulsion was also found to accelerate nucleation rates [25, 26]. Thus, nucleation in emulsions is far more complex and different mechanisms of heterogeneous nucleation can dominate crystallization rates.

Despite being a subject of considerable investigation, most nucleation studies interpret results based on individual droplet behavior. This is also evident from the underpinnings of the classical nucleation theory [16, 21, 27] which considers emulsion as an ensemble of independent stochastic nucleation sites in which nucleation may proceed through homogenous or heterogeneous mechanisms. It remains an open question whether droplet-droplet contacts can influence rates of nucleation in emulsions. It is important to address this question, not only because several nucleation studies use non-dilute emulsions where droplet crowding may have occurred, but also because concentrated emulsions are routinely employed in a variety of industrial products.

In this Letter, we study nucleation dynamics in a model concentrated emulsion - a hexagonally ordered monodisperse two-dimensional (2D) array of droplets. Monodispersity eliminates confounding effects of polydispersity and the hexagonal packing ensures uniform number of contacts between droplets. The 2D configuration allows direct observation of nucleation dynamics to correlate individual droplet nucleation events to system-wide effects. Our investigation reveals a collective mode of nucleation where solidified droplets nucleate neighboring droplets giving rise to an autocatalytic-like mechanism for accelerated crystallization in dense emulsions.

The model emulsion used is *n*-hexadecane-in-water – a popular alkane system used in nucleation studies [21, 24, 28] with a bulk melting point of $T_m$ = 18.2 °C [21]. The emulsion with < 5% polydispersity is made using microfluidics as described previously [29] and subsequently imbibed into a rectangular glass capillary of height $H$ and width $W$ such that $H \times W$ = 50 μm × 500 μm or 30 μm × 300 μm. During the capillary imbibition process, the confined droplets pack near the air-fluid interface creating a dense arrangement with a high degree of order and symmetry [30], as shown in Fig. 1a. Here, we study nucleation dynamics in 2D emulsions with droplet diameters $D$ = 24 μm or 40 μm stabilized with 2 wt% sodium dodecyl sulphate (SDS) or 2 wt% Tween 20. The 24 μm- and 40 μm- emulsion were confined in the 30 μm and 50 μm depth capillary respectively producing the same confinement of D/H = 0.8. The emulsion volume fractions could be varied from $\phi_v$ = 0.4 – 0.54, by tuning the packing density of droplets where $\phi_v = \frac{\pi}{12} \frac{ND^3}{WHL}$, $L$ = 2 mm is the length of the field-of-view and $N$ = 550 – 750 is the number of droplets in the field-of-view.

A typical experiment involves mounting a cut-and-sealed section of the glass capillary containing the droplet array on a Peltier-cooled thermal stage that has a temperature resolution of ± 0.01 °C. The emulsion was heated to a temperature T = 30 °C and subsequently cooled to 2 °C at a constant cooling rate of λ = 1 °C /min. During this linear cooling, the nucleation progression was imaged at a resolution of 2 µm/pixel (Supplementary Movie (SM) 1). We process the images to obtain both the position, solid fraction $n_S$, and statistical distributions of solid-drop cluster sizes based on Voronoi analysis [31-33]. Temperature variation across the capillary was considered negligible since the number density of solidified droplets in the left section of the capillary was not significantly different from the right section.

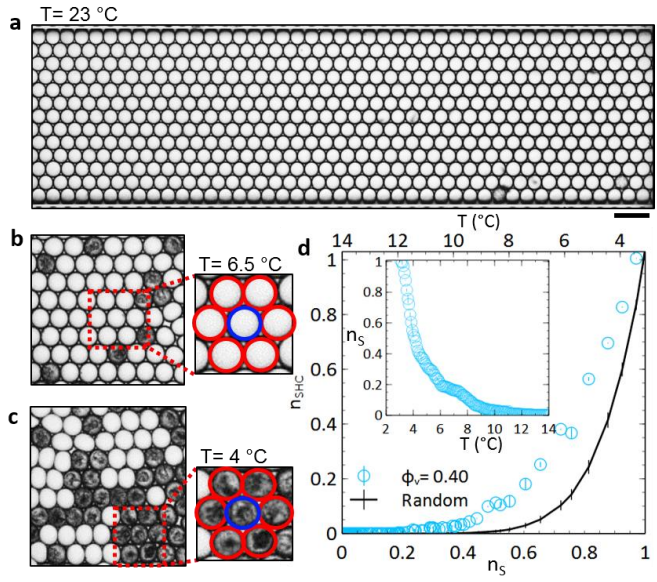

**Fig. 1.** (a) Image showing the 2D hexagonal droplet array for the study of collective nucleation dynamics. Scale bar is 100 µm. Images of nucleation in the droplet array at (b) 6.5, and (c) 4 °C. In (b) and (c), the insets indicate a liquid and a solid hexagonal cluster, respectively, where a central drop (blue) is surrounded by 6 immediate neighbors (red). (d) Number density of solid hexagonal clusters $n_{SHC}$ as a function of solid fraction $n_S$ in the array for SDS-stabilized emulsions of D = 40 µm and $\varphi_v = 0.4$. The solid line is the prediction from random nucleation simulation averaged over 100 runs. The inset in (d) shows the plot of the solid fraction vs. temperature. Error bar is standard deviation from two trials.

Fig.1 shows the representative data from a linear cooling of the 40 µm emulsion stabilized by 2wt% SDS. The emulsion remains as a system of supercooled liquid droplets (Fig. 1a) until the first nucleation events are observed at ∼14 °C, indicating an undercooling of ∼4 °C due to emulsification [19]. The images in Fig. 1b, c show a close view of the system at two temperatures highlighting nucleation propagation. The nucleated droplets appear darker, with their number density $n_S$ increasing as the temperature is lowered until all the droplets solidify at 3.5 °C (Fig.1d, inset). Additionally, we observe that the solidified droplets are non-spherical and the super-cooled liquid droplets are compacted. The compaction could be due to thermal contraction of emulsion droplets that is accompanied by unavoidable growth of nucleated bubbles (See SM 1).

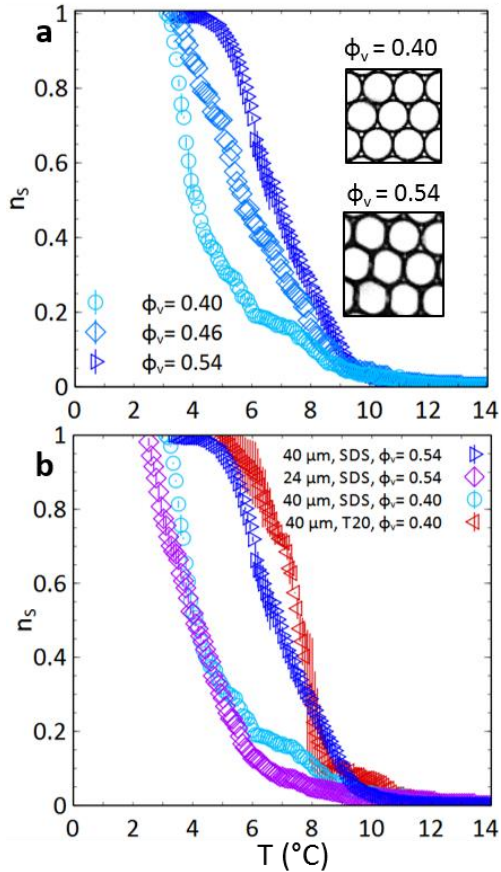

**Fig. 2.** Solid fraction as a function of temperature for (a) SDS-stabilized emulsions at volume fractions of 0.4, 0.46, and 0.54, D = 40 µm (b) emulsions with droplet sizes of 24 and 40 µm and surfactants of SDS and Tween 20. The insets in (a) show the droplet packing at volume fractions of 0.4 and 0.54. Error bar is standard deviation from two trials.

To assess whether the solidified droplets that are in direct contact with their supercooled neighbors influence their nucleation, we chose hexagonal clusters (Fig. 1b, inset), the repeating unit of the ordered array, and monitored them during cooling as they transform to solidified hexagonal clusters (SHCs, inset of Fig. 1c). Since each solidified droplet is in direct contact with 6-supercooled droplets, hexagonal clusters are a suitable choice for examining system-wide effects. In Fig. 1d, we show the number density of solidified hexagonal clusters $n_{SHC}$ in the array as a function of emulsion solid fraction as the system is cooled. We compared this data with that from a random nucleation simulation in which the droplet ensemble is chosen to be at a given solid fraction and individual droplets in the ensemble are assigned to nucleate randomly (SM 2).

Comparing the SHC results from the experiment and the random nucleation reveals two striking findings. First, the significant departure of the experimental data from the simulation shows that nucleation in the 2D emulsion does not proceed in a stochastic way as the classical nucleation theory suggests. Second, during bulk of the cooling process, the values of $n_{SHC}$ exceeds the random simulation result (Fig. 1d) indicating that solidified droplets in contact with their supercooled neighbors promote their nucleation.

The above results suggest a new collective mode of nucleation where, as the dense and ordered emulsion is cooled, a fraction of the supercooled droplets nucleate randomly. These 'random seeds' that are in direct contact with 6-supercooled neighbors solidify them probabilistically as the temperature is further lowered thereby effectively increasing the number of seeds available for further contact-driven nucleation of supercooled droplets. Such an autocatalytic-like mechanism significantly accelerates emulsion crystallization and requires sufficient thermodynamic driving force since the dramatic increase in SHCs occurs at $n_S > 0.3$ (Fig. 1d).

It is important to remark here that previously McClements and co-workers [26] have shown using ultrasound velocity measurements that addition of micron-sized solidified droplets to a bulk emulsion containing supercooled liquid droplets can accelerate the isothermal kinetics of nucleation. The enhancement in nucleation rate was hypothesized to be resulting from droplet collisions due to Brownian forces. Our contact-driven nucleation appears to be a non-Brownian analog with forced contacts due to the geometric constraints imposed in the ordered droplet array. The presence of interfacial crystals evident from the rough droplet surface [29] potentially act as sites for nucleation during contact. This contact-driven nucleation gives rise to the observed collective dynamics.

To gain insights into the collective mode of nucleation we altered system conditions by changing the volume fraction, droplet size, surfactant and cooling rate. With system conditions at $D = 40$ μm, 2 wt% SDS and $\lambda = 1°C/min$, we changed the volume fraction from 0.4 to 0.46 and 0.54, thereby increasing the droplet contact area, evident from the almost spherical shape at $\phi_v = 0.4$ to faceted polygons at $\phi_v = 0.54$. (inset of Fig. 2a). Data in Fig. 2a shows that as the temperature is lowered, initially the three emulsions have similar fraction of solidified droplets, however, with further cooling the solidification process is accelerated in the higher volume fraction emulsions. Thus, increasing the packing density or droplet contact area reduces the thermodynamic driving force for the autocatalytic mechanism to initiate. When subjecting the emulsion ($\phi_v = 0.40$) to different cooling rates ($\lambda$ = 0.2, 1.0 and 1.5 ºC/min), the entire emulsion solidification process remained unaltered (data not shown) suggesting that the collective mode of nucleation is more sensitive to the thermodynamic driving force than the rate of supercooling.

Maintaining the same surfactant and $\phi_v$ at 0.54 but decreasing the droplet size in the array from 40 µm to 24 µm reduced the solid fraction at a given temperature (Fig. 2b). This finding suggests that ordered arrays of smaller droplet size require more undercooling to trigger the collective mode of nucleation. When we changed from the ionic surfactant SDS to the non-ionic surfactant Tween 20, and tested at the lowest volume fraction of $\phi_v$ = 0.4, the degree of undercooling needed to initiate nucleation was not significantly different, but the auto-solidification process was dramatically accelerated (Fig. 2b) probably because the energy barrier for contact-driven nucleation is reduced when the surfactant has no charge [34].

Taken together, the results of Fig. 2 lend the following insights. A necessary criterion for the autocatalytic-like mechanism to trigger is an initial fraction of isolated 'random seeds'. The undercooling required to generate these random seeds appears to be dependent mostly on droplet size, and not so much on volume fraction, cooling rate and surfactant choice. This observation is consistent with the classical notion that small droplets have fewer impurities and therefore require greater undercooling to undergo homogeneous nucleation [20, 24]. Once the autocatalytic-like mechanism is engaged, the strength and nature of droplet contacts drive emulsion solidification. Stronger contacts due to compressive deformation of the droplet interface reduces the energetic barrier for propagating solidification. Likewise, the interfacial characteristics of the droplet can influence this energetic barrier.

Given our documentation of a novel collective mode of nucleation, next we sought to understand the dynamics of nucleation propagation, i.e. how does the system evolve from an initial fraction of random seeds to full solidification? Can we identify a framework that unifies the collective aspects of the nucleation dynamics despite system-specific differences? We speculated that since the hexagonally ordered array has 6-fold symmetry investigating the relationship between the number density of SHCs and solid fraction (c.f. Fig. 1d) might be a useful approach to track the evolution of the solidification process in the different systems we have studied.

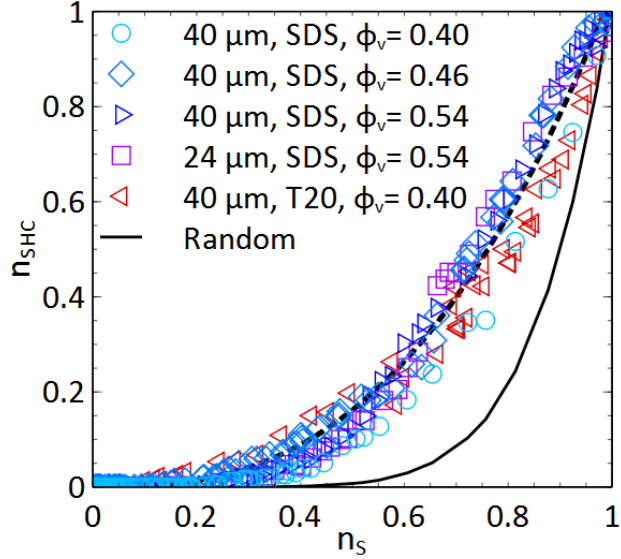

**Fig. 3.** Number density of solid hexagonal clusters as a function of solid fraction for emulsions with different droplet size, surfactant and volume fraction. The dashed and solid lines show the power law fit to the data and prediction from random nucleation simulation respectively.

Fig. 3 shows that the $n_{SHC}$ data from the systems studied nearly overlay each other despite differences in volume fraction, surfactant and droplet size. These data are still distinct from the random nucleation simulation results reinforcing the idea of solidified droplets promoting the nucleation of neighboring drops. A striking feature of this seemingly universal behavior is that it occurs without having to rescale any axis and can be captured with a power-law exponent of 2.68 ± 0.04. The theoretical basis of this power-law exponent is unclear, but it provides an empirical summary of our results.

The lack of a strong dependence of system conditions on $n_{SHC}$ vs. solid fraction signals the presence of a unifying mechanism that drives collective nucleation. However, $n_{SHC}$ only captures the solid hexagonal clusters, while during nucleation propagation we observe clusters with various droplet numbers (SM 1). To capture the statistical distribution of cluster sizes at different solid fractions, we tessellated the array by the Voronoi cells (Fig. 4a) surrounding each solid drop (SM 3). The Voronoi cell area A of each solid droplet shows the area in its nearest neighborhood that is covered by droplets that are all liquid except itself (Fig. 4a), dictated by the location and number of its nearest neighbor solid drops. For a unit SHC, the Voronoi cell is the hexagon surrounding the central drop ($A_0$) (Fig. 4b).

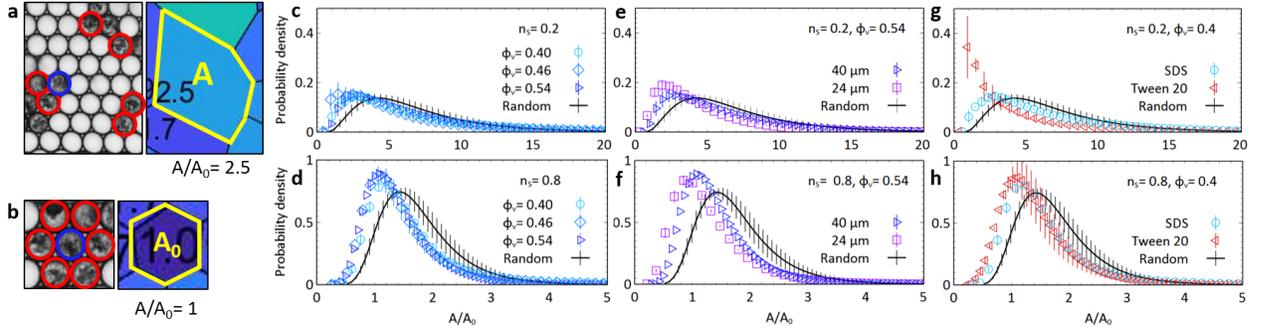

**Fig. 4**. (a) Left: image showing a candidate solid drop (blue) and its nearest neighbor solid droplets (red). Right: Voronoi area $A$ corresponding to the candidate solid droplet. (b) Voronoi area $A_0$ corresponding to a solid hexagonal cluster. The color scale corresponds to the variation in values of $A/A_0$ (see SM 3). The statistical distributions of normalized Voronoi area $A/A_0$ at (c, d) $n_S = 0.2$ and 0.8 for volume fractions of 0.4, 0.46, and 0.54 (e, f) $n_S = 0.2$ and 0.8 for droplet sizes of 40 and 24 µm and (g, h) $n_S = 0.2$ and 0.8 for emulsions stabilized with SDS and Tween 20. The solid line is the prediction from the random nucleation simulation. Error bar is standard deviation from two trials.

Fig. 4c-h show the statistical distributions of the Voronoi areas $A/A_0$ for arrays with solid fraction $n_S = 0.20$ and $0.80$. Surprisingly, the statistical distributions of $A/A_0$ for systems with different volume fractions (Fig. 4c,d) and droplet size (Fig. 4e,f) collapse onto a common distribution except for the Tween 20 system at low solids fraction. As expected, these distributions are distinct from the random simulation curves. Thus, as the nucleation proceeds in the droplet array, in general we find the statistical distribution of cluster sizes remains invariant. In the case of Tween 20 system, we find that at low solid fraction, $n_S = 0.20$, the distribution peaks at $A/A_0 \approx 1$ suggesting that several SHCs are formed early which then propagate nucleation to arrive at the invariant distribution found in other systems at high solid fraction.

What does the collapse of data in Fig. 3 and the invariant distributions found in Fig. 4 suggest about the dynamics of nucleation propagation? It is possible that in our geometrically constrained ordered array, during cooling, the ensemble of droplets experiences thermal contraction allowing mechanical stress to propagate between droplets promoting contact-driven nucleation. Due to hexagonal symmetry, our system has 6 degrees of freedom lending sufficient flexibility for the stress to propagate and yielding statistical distribution of cluster sizes that are not sensitive to system-specific details.

In summary, we report a new collective mode of nucleation that occurs in dense emulsions where crowding is inevitable. This new mechanism is in striking contrast to previous works that discuss nucleation mechanisms based on individual droplet behavior. We show that statistical distributions of solidified droplet clusters during nucleation propagation are independent of emulsion characteristics except surfactant. Moving beyond the classical theory of nucleation, our work motivates the need for a new theoretical description of emulsion crystallization that considers collective effects. In the broader context, our study is a novel addition to the growing literature that report fascinating collective phenomena exhibited by densely-packed microfluidic emulsions [32, 35].

The authors thank the financial support of the Jack Maddox Foundation and National Science Foundation (CAREER: 1150836). We are grateful to Jose Alvarado, Amir Hajiakbari, Hamed Sari-Sarraf, Gordon Christopher and Brandon Weeks for stimulating discussions. We also thank Sindy Tang and Ya Gai for sharing their image processing code for droplet-tracking.